\documentstyle[preprint,aps]{revtex}

\begin{document}
\draft
\title{The Effects of Resonant Tunneling 
on Magnetoresistance through a Quantum Dot}
\author{Tetsufumi Tanamoto and Shinobu Fujita}
\address{Research and Development Center, 
Toshiba Corporation, 
Saiwai-ku, Kawasaki 210-8582, Japan}
\date{\today}
\maketitle
\begin{abstract}
The effect of resonant tunneling  
on magnetoresistance (MR) is studied theoretically 
in a double junction system. 
We have found that the ratio of the MR of the resonant peak current
is reduced more than that of the single junction,  
whereas that of the valley current is 
enhanced depending on the change of the discrete energy-level 
under the change of magnetic field.  
We also found that the peak current-valley current (PV) ratio 
decreases when the junction conductance increases. 
\end{abstract}

\medskip

\narrowtext
\section{Introduction}
As nano-fabrication technology using magnetic materials advances,  
the magnetoresistance (MR) in mesoscopic systems 
has attracted growing interest, mainly because of 
its possible applications in storage devices\cite{Johnson,Moodera}. 
The resonant tunneling phenomenon is one of the events 
expected to occur in magnetic nano-particles or thin film systems 
where discrete energy-level structures are prominent.  
Ferromagnet/insulator/ferromagnet/insulator/ferromagnet 
(FM/I/FM/I/FM) is 
a basic structure of this system. 
The electron tunneling through the insulator 
is considered to be based on the independent polarized electrons  
and the difference in MR is considered to be derived from 
the difference of the spin polarized  free electron 
tunneling\cite{Maekawa,Slonczewski}. 
Recently, Zhang {\it et al.}\cite{Zhang} calculated the $I$-$V$ 
characteristics of the FM/I/FM/I/FM structure based 
on the Tsu-Esaki formulation 
(S-matrix theory) and showed that the MR is more than 90\%, 
which is a great enhancement compared with the case of a single junction
. 
However, the origin of the enhancement is not clear. 
In addition, although the peak current-valley current (PV) ratio 
is one of the key factors of resonant tunneling, the effect of the 
MR on this ratio has not been clarified. 
This is because S-matrix theory is not easily applied to  
 analyze the underlying physics, in spite of 
its usefulness as a more realistic method  
for self-consistent calculations.  
In this paper we discuss the effects of the resonant tunneling
on the MR {\it analytically}, based on the two band spin polarized free 
electron model. 
We assume that the capacitances of two junctions are large and 
neglect charging effects.  
The general formula of current-voltage characteristics is 
obtained by elaborating that derived by  
Jauho {\it et al.}\cite{Jauho} 
and is shown to be a useful formula for investigating the 
detailed physics of a double barrier structure. 
As we are using Green's function method, 
our model can be extended, by refining 
the self-energy part of Green functions, 
to be applied to the case 
in which free electron approximation cannot be used 
because of the existence of some scatterings,  
although it is not the subject of this paper.  

\section{Formulation of the Description of the System}
The Hamiltonian consists of the electronic part $\hat{H}_{\rm E}$, 
and the transfer part $\hat{H}_{\rm T}$. 
The electronic part consists of 
electrode and island parts, 
\begin{equation}
\hat{H}_{\rm E} =  \! \! \! \!
\sum_{{\bf k}, \alpha \in L,R, \sigma }  \! \! \! \! 
E_{{\bf k}\alpha \sigma} \hat{c}_{{\bf k}\alpha \sigma }^\dagger 
\hat{c}_{{\bf k}\alpha \sigma} \! +\!
\sum_{m \sigma} E_{m \sigma } 
\hat{d}_{m \sigma}^\dagger \hat{d}_{m \sigma}, 
\end{equation}
where $\alpha$ represents a set of parameters that, together 
with wave vectors ${\bf k}$, completely specify the 
electronic state of the left (L) or right (R) electrode, 
and $m$ specifies the energy levels of the central island.
By an internal magnetic field ${\bf h_\alpha}$ and 
with the Pauli spin matrix ${\bf \sigma}$, the energy dispersion 
relation is expressed as\cite{Slonczewski,Zhang} 
\begin{equation}
E_{{\bf k}\alpha {\bf \sigma}}=
\hbar^2 {\bf k}_{\bf \sigma}^2 /(2m)
-{\bf h}_\alpha \cdot {\bf  \sigma}. 
\label{dispersion}
\end{equation} 
Hereafter we write 
$E_{{\bf k}\alpha \uparrow} = 
\hbar^2 {\bf k}_\uparrow^2/(2m) - h_\alpha$ 
and 
$E_{{\bf k}\alpha \downarrow} =
\hbar^2 {\bf k}_\downarrow^2 /(2m) + h_\alpha$. 
The transfer part is described by
\begin{equation}
\hat{H}_{\rm T} = \! \! \sum_{n{\bf k} \alpha \in L,R \sigma} 
\! \! [ V_{n {\bf k}\alpha \sigma} (t)
\hat{c}_{{\bf k}\alpha \sigma}^\dagger  \hat{d}_{n \sigma} 
+ \mbox{h.c.} ].
\end{equation}

The current at junction $\alpha$ (=L or R) 
is given by\cite{Jauho,Mier}
\begin{eqnarray}
J_\alpha (t) 
&=& (-)^{\beta} e \sum_{\sigma} 
\langle \dot{\hat{N}}_{\alpha \sigma} \rangle 
= -(-)^{\beta}\frac{ie}{\hbar} \sum_{\sigma} 
\langle [\hat{H}, \hat{N}_{\alpha \sigma}] \rangle 
\nonumber \\
&=& (-)^{\beta}\frac{ie}{\hbar} \sum_{n{\bf k} \sigma} 
[V_{n {\bf k}\alpha \sigma} 
\langle \hat{c}_{{\bf k}\alpha \sigma}^\dagger  
\hat{d}_{n \sigma}  \rangle 
-V_{n {\bf k}\alpha \sigma}^* \langle 
\hat{d}_{n \sigma}^\dagger  \hat{c}_{{\bf k}\alpha \sigma} 
 \rangle ]
\nonumber \\
&=& (-)^{\beta}\frac{2e}{\hbar} {\rm Re} 
\left\{ \sum_{{\bf k}n \sigma} 
V_{n {\bf k}\alpha \sigma} (t) 
G^{<}_{n{\bf k} \alpha \sigma} (t,t) \right\},
\label{eqn:J0}
\end{eqnarray}
where $\beta$=0 for the left junction and $\beta$=1 
for the right junction 
and $G_{n{\bf k}\alpha \sigma}^< (t,t')  \equiv  
i\langle \hat{c}_{{\bf k}\alpha \sigma}^\dagger (t) 
\hat{d}_{n \sigma} (t') \rangle $
is an analytic continuation 
of the contour-ordered Green's function 
$G^{<}_{n{\bf k} \alpha \sigma} 
(\tau, \tau')$ which is defined
in the interaction representation by
\begin{eqnarray}
G_{n{\bf k}\alpha \sigma} (\tau, \tau')  
&\equiv&  i\langle T_C \{ 
\hat{c}_{{\bf k}\alpha \sigma}^\dagger (\tau') 
\hat{d}_{n \sigma} (\tau) \nonumber \\
&\times & \! \exp \left(-\frac{i}{\hbar} \! \! \int_C \! 
\hat{H}_{\rm T} (\tau_1) d \tau_1 \right) \} \rangle, 
\label{eqn:ggtau}
\end{eqnarray}
where $T_C$ is the contour-ordering operator. 
We assume that electrons in the left and right electrodes
are noninteracting. 
Then the only nonvanishing terms in Eq.\ (\ref{eqn:ggtau}) are 
those in which $\hat{c}_{{\bf k}\alpha}^\dagger (\tau')$ 
is contracted with 
$\hat{c}_{{\bf k}\alpha} (\tau_1)$ in the exponential term. 
We then obtain, after analytic continuation to real time, 
\begin{eqnarray}
G^{<}_{n{\bf k}\alpha \sigma} (t,t') 
&=& \! \! \sum_m \int_{-\infty}^{\infty} \! \! \! \frac{dt_1}{\hbar}
 V^*_{n{\bf k}\alpha \sigma} (t_1) 
[ G_{nm \sigma}^r (t,t_1) 
g_{{\bf k}\alpha \sigma}^< (t_1, t')  \nonumber \\
& & + G_{nm \sigma}^< (t,t_1) 
g_{{\bf k}\alpha \sigma}^a (t_1, t') ],
\label{eqn:Gd}
\end{eqnarray}
where $G^{<}_{nm \sigma}(t,t') \equiv i \langle 
\hat{d}_{m \sigma}^\dagger (t') \hat{d}_{n \sigma} (t) \rangle$ 
is the Green's function of the central island,    
and 
\begin{eqnarray}
g^{<}_{{\bf k}\alpha \sigma}(t_1, t_2) &\equiv& 
i \langle \hat{c}_{{\bf k} \alpha \sigma}^\dagger (t_2) 
\hat{c}_{{\bf k} \alpha} (t_1) \rangle , \\ 
g^{>}_{{\bf k}\alpha \sigma}(t_1, t_2) &\equiv& 
-i \langle \hat{c}_{{\bf k} \alpha \sigma} (t_1) 
\hat{c}_{{\bf k} \alpha \sigma}^\dagger (t_2) \rangle .
\end{eqnarray}
Substituting Eq. (\ref{eqn:Gd}) into Eq. (\ref{eqn:J0}), 
we obtain 
\begin{eqnarray}
\lefteqn{ J_\alpha (t) \!  =(-1)^{\beta} \frac{e}{\hbar^2} \! 
\sum_{{\bf k} mn \sigma} \! V_{n{\bf k} \alpha \sigma}(t) \!
\int_{-\infty}^{\infty}  \! \! \! \! \!
dt' V^*_{m{\bf k} \alpha \sigma} (t') 
\{ G^>_{nm \sigma} (t,t') } \nonumber \\  
& & \times  
g^<_{{\bf k} \alpha \sigma} (t',t)  
-  G^<_{nm \sigma} (t,t') g^>_{{\bf k} \alpha \sigma} (t',t) \} ,
\label{eqn:jstrt}  
\end{eqnarray}
where $G^>_{nm \sigma}(t,t')\equiv 
-i\langle 
\hat{d}_{n \sigma} (t) \hat{d}_{m \sigma}^\dagger (t')
\rangle$.
By introducing a noninteracting self-energy, 
$\Sigma_0^{><}$, defined 
by $G_0^{><} = G^r_0 \Sigma_0^{><} G^a_0$, 
and using the Dyson equations, $(1+G^r \Sigma^r)G^r_0=G^r$ 
and $G^a_0 (1+\Sigma^a G^a)=G^a$,
we may cast the Keldysh equation
$G^{><} =(1+G^r \Sigma^r)G_0^{><} (1+\Sigma^a G^a) 
+ G^r \Sigma G^a$ into the following form  \cite{Sarker}:
\begin{eqnarray}
G_{nm \sigma}^{><} (t,t') 
&=& \! \!  \sum_{n_1,n_2 }
\! \int \! dt_1 dt_2 G^r_{nn_1 \sigma} (t,t_1) 
\Sigma^{><}_{{\rm tot} \ n_1n_2 \sigma} (t_1,t_2) 
\nonumber \\
&\times &G^a_{n_2 m \sigma} (t_2,t') ,  
\label{Keldysh eq}
\end{eqnarray}
where
\begin{eqnarray}
\Sigma^{><}_{{\rm tot} \ n_1n_2 \sigma} (t_1,t_2)
&=& \Sigma^{><}_{0 \ n_1n_2 \sigma} (t_1,t_2)
+ \Sigma^{><}_{{\rm T} \ n_1n_2 \sigma} (t_1,t_2) 
\label{eqn:eta0}
\end{eqnarray}
The first term on the right-hand side describes the self-energy
of the island in the absence of disorder, interaction, and tunneling. 
Since it corresponds to the free part, it is infinitesimal:
$\Sigma^{><}_{0n_1 n_2 \sigma} (\epsilon)= 2i \delta_{n_1 n_2} \eta 
[f_0 (\epsilon)-1/2 \mp 1/2]$, 
where $\eta$ is a positive infinitesimal, $>(<)$ refers to the 
minus (plus) sign,
$f_0(\epsilon) = (\exp \beta (\epsilon-V_d) +1 )^{-1}$, 
and  $V_d$ is the bottom of the island energy.
The second term describes a self-energy due to tunneling:
\begin{equation}
\Sigma^{><}_{{\rm T} n_1n_2 \sigma} (t_1,t_2)
=\sum_{{\bf k} \alpha} 
\frac{V_{n_1 {\bf k} \alpha \sigma}^*(t_1) 
V_{n_2 {\bf k} \alpha \sigma}(t_2) }{\hbar^2}
g_{{\bf k} \alpha \sigma}^{><} (t_1,t_2).
\label{eqn:sigma}
\end{equation}
Because of the infinitesimal factor $\eta$,
the free part is  important only when the remaining parts 
are absent (it is not the present case).

In the following we focus on the case in which 
$\Gamma_{nn'{\bf k}\alpha \sigma} (t,t')$  $\equiv$ 
$2\pi$ $V_{n{\bf k} \alpha \sigma}^*(t)$  
$V_{n'{\bf k} \alpha \sigma}(t')$ 
is a real function of $t-t'$.  
From Eq.\ (\ref{eqn:jstrt}), $J_L$-$J_R$ =$ e\sum_\sigma 
\int_{-\infty}^{\infty} 
dt' $ $(G_{nm \sigma}^{>}(t,t') \Sigma_{{\rm T}mn \sigma}^{<}(t',t)$
$-$ $G_{nm \sigma}^{<}(t,t') \Sigma_{{\rm T}mn \sigma}^{>}(t',t))$. 
Because $G_{nm \sigma}^{><}(t,t')$ contains 
$\Sigma_{{\rm T}mn \sigma}^{><}(t,t')$
(Eq. (\ref{eqn:sigma})), 
the conservation of current through the two 
junctions is automatically satisfied, that is, $J_L=J_R$.
Thus we can discuss the current through the junctions 
by $J_L$.

The retarded and advanced Fourier-transformed 
Green's functions at the central island, 
$G^r_{nn' \sigma} (\epsilon)$ and $G^a_{nn'\sigma} (\epsilon)$, 
are derived from the Dyson equation:
\begin{eqnarray}
\lefteqn{\hbar [G^{r,a}_{nn'\sigma} (\epsilon)]^{-1} 
= \hbar [ g^{r,a}_{n \sigma} (\epsilon) ]^{-1} 
- \hbar \Sigma^{r,a}_{{\rm tot} \ nn'\sigma} (\epsilon) }
\nonumber \\ 
&=& \! \epsilon \! -\! (E_{n\sigma} \! +\! V_{d\sigma}) 
\! - \! \Lambda_{nn'\sigma} (\epsilon) \! \pm \! 
\frac{i}{2} \left( 2\eta  
\! + \! \Gamma_{nn'\sigma} (\epsilon) \right),
\end{eqnarray} 
where 
$\Gamma_{nn'\sigma} (\epsilon) \equiv 2 {\rm Im} 
\Sigma^{a}_{{\rm T} nn'\sigma} (\epsilon)$, 
and $2\pi \hbar \Sigma^{r,a}_{{\rm T} nn'\sigma} (\epsilon) \! \! 
= \! \! \sum_{{\bf k} \alpha } 
\Gamma_{nn'{\bf k} \alpha \sigma} (\epsilon) 
g_{{\bf k} \alpha \sigma}^{r,a} (\epsilon)$.  
The real part of the self-energy 
$\Lambda_{nn'\sigma} (\epsilon)$ shifts energy-levels
in the central island and we will regard this effect as included 
in our assumed one-body energy levels of the quantum dot. 
Here $\Gamma_{nn'\sigma}(\epsilon)$ 
shows the half-width of resonant peaks :
\begin{equation}
\Gamma_{nn'\sigma} (\epsilon) 
= \sum_{{\bf k} \alpha} 
\! \int \! d \epsilon_1 
\Gamma_{nn'{\bf k} \alpha} (\epsilon_1) 
\delta ( \epsilon-\epsilon_1 -E_{{\bf k} \alpha \sigma} ) 
\label{eqn:GLR}
\end{equation}
With  $ A_{n_1 n_2 \sigma}(\epsilon) 
\equiv i (G^r_{n_1 n_2 \sigma}(\epsilon) 
-G^a_{n_1 n_2 \sigma} (\epsilon))$, 
and $\hbar \Theta_{nm \sigma} (\epsilon) \equiv 
G^r_{nn_1 \sigma} (\epsilon) G^a_{n_2m \sigma} (\epsilon) /  
 A_{n_1 n_2 \sigma} (\epsilon) $,  
$J_L$ can be cast into the following form,  
\begin{eqnarray}
\lefteqn{J_L \! =\!  \frac{e}{2\pi \hbar^2} 
 \! \! \! \sum_{
\stackrel{\scriptstyle {\bf kk'} nm}{{n_1 n_2 \sigma}}
} \! \!   
\int_{-\infty}^{\infty} \! \! \! \! \! \! d \epsilon  
\Gamma_{nm{\bf k}L \sigma}^* (\epsilon)
\Gamma_{n_1 n_2 {\bf k'} R \sigma}^* (\epsilon)
A_{n_1 n_2 \sigma} (\epsilon) 
 }\nonumber  \\ 
& & \times \Theta_{nm \sigma} (\epsilon) 
\{ f_{L} (E_{{\bf k}L \sigma}\! ) \! 
-\!  f_{R}(E_{{\bf k}'R \sigma} \!) \}
\delta (\epsilon \!- \! E_{{\bf k}L\sigma}\! ) 
\delta (\epsilon \!- \! E_{{\bf k}'R\sigma}\! ).
\nonumber  \\
\label{eqn:mr} 
\end{eqnarray}
This is a general expression of current 
where $f_L(\epsilon) \! \! =\! \! 1/ \! 
( e^{\beta(\epsilon  -\! E_{F \sigma} \! -\! eV)} \! 
\! +\! 1 )$ and $f_R(\epsilon) \! \! = \! \! 
1/ \! ( e^{\beta(\epsilon -E_{F \sigma})} \!  \! +\!  1 ) $ with 
$E_{F \sigma}$ being the Fermi energy of the electrodes. 
Hereafter, we apply this expression to simple cases. 
First we assume that $\Sigma^{r,a}_{mn\sigma} (\epsilon) =
\delta_{mn} \Sigma^{r,a \sigma}_n$.   
This corresponds to a situation in which 
energy levels in the island are mutually uncorrelated 
during the tunneling process. 
Then all Green's functions are diagonalized and  
$A_{n_1 n_2 \sigma} (\epsilon)$ reduces to 
\begin{equation} 
A_{n \sigma} (\epsilon)= \frac{\hbar \Gamma_{n \sigma} (\epsilon) }
{\left[ \epsilon-( E_{n \sigma} +V_{d \sigma} )  
- \Lambda_{n \sigma} (\epsilon) \right]^2 
+\left[ \Gamma_{n \sigma} (\epsilon) \right]^2 /4 }.  
\label{an}
\end{equation} 
and $\Theta_{n \sigma} (\epsilon)$ reduces to 
$\Gamma_{n \sigma} (\epsilon)^{-1}$. 

$\Gamma_{nn'\sigma} (\epsilon)$ is related with 
the density of state (DOS) of the electrode such that 
\begin{eqnarray}
\Gamma_{nn'\alpha \sigma} (\epsilon)
&=&\sum_{{\bf k}} V_{n{\bf k}\alpha \sigma}^*
V_{n'{\bf k}\alpha \sigma} 
\delta (\epsilon-E_{{\bf k} \alpha \sigma}) \nonumber \\
&\approx& { {\cal V}} \int \frac{4\pi k^2 d k}{(2\pi)^3} 
V_{n{\bf k}_F\alpha \sigma}^* V_{n'{\bf k}_F\alpha \sigma} 
\delta (\epsilon-E_{{\bf k} \alpha \sigma}) \nonumber \\
&=& 
V_{n{\bf k}_F\alpha \sigma}^* V_{n'{\bf k}_F\alpha \sigma} 
D_{\alpha \sigma} (\epsilon), 
\label{gamma_h}
\end{eqnarray}
where $D_{\alpha \sigma} (E)$ is a density of state 
expressed by $D_{\alpha \sigma} (E)$
= $\frac{{\cal V}}{4\pi^2} \left( \frac{2m}{\hbar^2}\right)^{3/2}
\sqrt{E+\sigma h_\alpha} $ in a three-dimensional space. 
This approximation will be most suitable when the interfacial 
material is a quantum box or particles.
From the relation  
$\Gamma_{\alpha \sigma} (\epsilon)= 2 \pi 
D_{\alpha \sigma} (\epsilon)
|V_{{\bf k}\alpha \sigma}|^2$,
resistances $R_{\alpha \sigma} (\alpha=L, R)$ can be evaluated 
to be $ R_{\alpha \sigma} /R_K = (\Gamma_{\alpha \sigma}(E_{F \sigma}) 
D_d (E_{F \sigma}))^{-1}$,
where $R_K$ is the resistance quantum 
$h/e^2$ =25.8k$\Omega$ and $D_d (E_{F \sigma})$ is 
the DOS of the island  at the Fermi energy. 
In this paper MR is discussed in terms of the change of 
$\Gamma_{\alpha \sigma}$ instead of $R_{\alpha \sigma}$

We see $\Gamma_{n \alpha \sigma}(E_{{\bf k}\alpha \sigma})$ 
as a function of the internal magnetic field :
$\Gamma_{n\alpha \sigma}(E_{{\bf k}\alpha \sigma}) 
\equiv \Gamma_{n\alpha \sigma}(h_\alpha)$. 
Thus we have the resonant tunneling formula \cite{Weil}
under an external magnetic field, $H$, as,
\begin{equation}
J_{L}(H) = \frac{e}{2\pi \hbar} 
\! \sum_{n \sigma}  \int_{s_0}^{\infty} 
\! \! \! \! \! \! 
d E_{{\bf k}L \sigma}
\frac{\Gamma_{nL \sigma}(h_L) \Gamma_{nR \sigma}(h_R)}
{\left[ E_{{\bf k}L \sigma}-( E_{n \sigma} +V_{d \sigma} )  
 \right]^2 +\left[
\frac{ \Gamma_{n L \sigma}(h_L) + 
\Gamma_{n R \sigma} (h_R)}{2}\right]^2 } 
\{ f_{L} (E_{{\bf k}L\sigma}) - f_{R}(E_{{\bf k}L\sigma}) \},  
\label{ivrtc}
\end{equation}
where $s_0={\rm max}(eV\! -\! \sigma h_L, -\! \sigma h_R)$. 

Here we compare the MR of the resonant tunneling 
current with that of the single junction by their conductances.  
The conductance of the system is given by 
$G_{\sigma} \equiv \partial J_{\sigma}/\partial V$, 
and the magnetoresistance is defined as 
\begin{eqnarray} 
MR &\equiv& \frac{G_{\uparrow \uparrow}-
G_{\uparrow \downarrow}}{G_{\uparrow \downarrow}} \nonumber \\
&=&\frac{\sum_\sigma (G_\sigma (H)-G_\sigma (-H))}
{\sum_\sigma G_\sigma (-H)}. 
\end{eqnarray}
We have a picture that the change of magnetic field 
makes the distribution of the DOS (Fig.\ref{fig1}) and the 
$\Gamma_{\alpha \sigma} (h_\alpha)$ (Eq.(\ref{gamma_h})).
We consider the derivation of 
$\Gamma_{\alpha \sigma} (h_\alpha)$ 
from the case where the internal 
magnetic field is zero, $h_\alpha =0$: 
\begin{eqnarray}
\Gamma_{nL\sigma}(h_L) &=& \Gamma_{nL\sigma}(0)
(1+\Delta_{L\sigma} (h_L)), \nonumber \\
\Gamma_{nR\sigma}(h_R) &=& \Gamma_{nR\sigma}(0)
(1+\Delta_{R\sigma} (h_R)),  \nonumber \\ 
E_{n \sigma}(h_n) &=& E_{n \sigma}(0) 
(1+\gamma_\sigma (h_n)),  
\label{Gamma_h} 
\end{eqnarray} 
where $| \Delta_{\alpha \sigma} (h_\alpha) | \ll 1$ 
($\alpha$=$L$, $R$) and  
 $| \gamma_\sigma (h_n) | \ll 1$ and we set   
$ \Gamma_{n\alpha \sigma} (0)=
\Gamma_{n\alpha}$ and $E_{n \sigma}(0)=E_n$ 
in the following. 
Note that  $\gamma_\sigma (h_n)$ depends on the relative 
displacement of the energy-level compared with the two 
electrodes.  
The spin polarized current through the single junction, 
$J_{\sigma}^{(S)}(H) $ at $\alpha$=$L$ is given from 
Eq.(\ref{eqn:jstrt}) with  
$n,m \rightarrow {\bf k'}_R$  and 
$G_{nm \sigma}^{><} (t) \rightarrow g_{{\bf k'}_R \sigma}^{><}(t)$ as 
\begin{eqnarray}
\lefteqn{J_{\sigma}^{(S)}(H)=\frac{e}{2\pi \hbar} 
\! \int_{s_0}^{\infty} 
\! \!
d E_{{\bf k}L \sigma}
\Gamma_{LR \sigma}(h_L) D_{R\sigma}(h_R) 
}\nonumber  \\
& & \times
\{ f_{L} (E_{{\bf k}L\sigma}) - f_{R}(E_{{\bf k}L\sigma}) \}, 
\label{ivsj}
\end{eqnarray}
and the conductance $G_{\sigma}^{(S)}(H)$ is expressed as 
$G_{\sigma}^{(S)}(H) \equiv 
\Gamma_{L\sigma}(h_L) D_{R\sigma}(h_R)$ 
$(\sim |V_{\sigma{\bf k}_F}|^2 D_{L\sigma}(h_L) D_{R\sigma}(h_R))$. 
We consider the case where the left electrode is 
a soft magnet and the right one is a hard magnet. 
Then $\Delta_{R\sigma}(h_R)$ does not change 
under the inversion of the external magnetic field $H$,    
and  the MR of the single junction 
is given by
\begin{equation}
MR^{(S)}=\sum_{\sigma}\frac{1}{2}(\Delta_{L\sigma} (h_L) 
-\Delta_{L\sigma} (-h_L) ).
\label{mr_sj}
\end{equation}
Next we derive the MR of the double barrier structure. 
Conductance $G_{n \sigma}$ near the $n$-th energy-level 
in the island is given by  
using $\partial f(\epsilon)/\partial \epsilon 
=-\delta (\epsilon -\mu)$ 
($T \! \rightarrow \! 0$) in Eq.(\ref{ivrtc}) :  
\begin{equation}
G_{n \sigma}(H) \approx  \frac{e^2}{2\pi \hbar}   
\frac{\Gamma_{nL \sigma}(h_L) \Gamma_{nR \sigma}(h_R)}
{\left[ E_F- E_{n \sigma} (h_n) \right]^2 +\left[
\frac{ \Gamma_{n L \sigma}(h_L) 
+ \Gamma_{n R \sigma} (h_R)}{2}\right]^2 }. 
\end{equation}
From this we obtain the conductance at the peak current 
(current at resonance) 
$G_{n \sigma}^{\rm res} (H) $ and that at the valley current 
(current at off-resonance )
$G_{n \sigma}^{\rm off} (H) $:
\begin{eqnarray}
G_{n \sigma}^{\rm res} (H) \! &\approx & \! \frac{2e^2}{\pi \hbar}   
\frac{\Gamma_{nL \sigma}(h_L) \Gamma_{nR \sigma}(h_R)}
{\left[ \Gamma_{n L \sigma}(h_L) 
+ \Gamma_{n R \sigma} (h_R) \right]^2 },  \label{g_res}\\
G_{n \sigma}^{\rm off} (H) \! &\approx & \! \frac{e^2}{2\pi \hbar}   
\frac{\Gamma_{nL \sigma}(h_L) \Gamma_{nR \sigma}(h_R)}
{[E_F- E_{n \sigma} (h_n)]^2}. \label{g_off}
\end{eqnarray} 
Here we assume that $E_F- E_{n \sigma} (h_n) >0 $.
The change of the conductances 
Eq.(\ref{g_res}) and  Eq.(\ref{g_off}) 
are given in order of $\Delta_{L\sigma} (h_L)$ as
\begin{eqnarray}
& & G_{n \sigma}^{\rm res} (H) \! 
-\! G_{n \sigma}^{\rm res (0)}  \nonumber \\
&\approx & \! \! \sum_{\sigma} \! \frac{2e^2}{\pi \hbar}   
\frac{\Gamma_{nL} \Gamma_{nR}}
{\left[ \Gamma_{nL}  \! \! + \! \! \Gamma_{nR} \right]^3 } 
(\Gamma_{nR} \! \!  -\! \! \Gamma_{nL})
\Delta_{L\sigma} (h_L) , \\
& & G_{n \sigma}^{\rm off} (H) \! 
-\! G_{n \sigma}^{\rm off (0)}  \nonumber \\
&\approx & \! \! \sum_{\sigma} \! \frac{e^2}{2\pi \hbar}   
\frac{\Gamma_{nL} \Gamma_{nR}}
{[E_F\! \! -\! \!  E_{n \sigma}]^2} 
\left[ \Delta_{L\sigma} (h_L) \! \!+ \! \!
\frac{2E_{n \sigma}}{E_F \! \! 
-\! \! E_{n \sigma}}\gamma_\sigma (h_n) \right].
\end{eqnarray}
where $G_{n \sigma}^{\rm res (0)}$ and 
$G_{n \sigma}^{\rm off (0)}$ are values when there is 
no internal magnetic field. 

Thus MR of the resonant current and off-resonant 
current  are given as 
\begin{eqnarray}
M \! R_{n}^{{\rm res}} \! \! &=& \! \! \sum_\sigma \frac{1}{2} 
\frac{\Gamma_{nR}\! -\! \Gamma_{nL}}{\Gamma_{nR}\! +\! \Gamma_{nL}}
(\Delta_{L\sigma} (h_L) \!-\! \Delta_{L\sigma} (-h_L) ), 
\label{MR_res}\\
M \! R_{n}^{{\rm off}} \! \! &=& \! \! \sum_{\sigma} \! 
\frac{1}{2} [ \Delta_{L\sigma} (h_L) \!- \! 
\Delta_{L\sigma} (-\! h_L) 
\nonumber \\
& +& \! \frac{2E_{n \sigma}}{E_F \! -\! E_{n \sigma}\! }
(\gamma_\sigma (h_n) \!-\! \gamma_\sigma (- \! h_n )) ] . 
\label{MR_off}
\end{eqnarray}
Because of the factor $(\Gamma_{nR}-\Gamma_{nL})/
(\Gamma_{nR}+\Gamma_{nL})
< 1$, 
Eq.(\ref{MR_res}) shows that the ratio of 
MR of the peak current is smaller than that 
of the single junction (Eq.(\ref{mr_sj})), whereas  
 Eq.(\ref{MR_off}) shows that the MR of 
the valley current is enhanced depending on 
the change of the energy-level in the island 
({\it i.e.} if $\gamma_\sigma (h_n) $ - 
$\gamma_\sigma (- \! h_n )$ has the same sign as 
$\Delta_{L\sigma} (h_L)$ - 
$\Delta_{L\sigma} (-\! h_L)$). 
In the later numerical calculations, the enhancement of 
the valley current is shown 
as the simplest case where the island is 
non-magnetic material. 
Thus the MR enhancement of more than 90\% of 
the double junction shown by Zhang\cite{Zhang} 
is found to be the enhancement of the valley current 
and NOT that of the peak current. 
From these results, it is easy to conjecture 
the PV ratio, defined 
$PV_n(H)$ $\equiv$ $\sum_\sigma 
G_{n \sigma}^{\rm res}(H)$ /$
\sum_\sigma G_{n \sigma}^{\rm off}(H)$, decreases. 
The explicit expression of 
the PV ratio in changing the external magnetic 
field is given as  
\begin{eqnarray}
\lefteqn{PV_n (H)-PV_n (-H) = 
\frac{\sum_\sigma G_{n \sigma}^{\rm res}(H)}
{\sum_\sigma G_{n \sigma}^{\rm off}(H)}
-\frac{\sum_\sigma G_{n \sigma}^{\rm res}(-H)}
{\sum_\sigma G_{n \sigma}^{\rm off}(-H)}  }\nonumber \\
&\approx& -PV_n(0) 
\sum_\sigma ( 
\frac{\Gamma_{nL}}{\Gamma_{nR} \! + \! \Gamma_{nL}}
(\Delta_{L\sigma} (h_L) \! - \! \Delta_{L\sigma} (-h_L))  
\nonumber \\
& +&\! \frac{E_{n \sigma}}{E_F \! -\! E_{n \sigma}}
(\gamma_\sigma (h_n) -\gamma_\sigma (-h_n) )). 
\end{eqnarray}
This shows that PV ratio is reduced 
when the junction conductance increases under the  
change of the direction of the external magnetic field. 

Although these results are derived starting from 
Eq. (\ref{Gamma_h}), they are also valid for a one band model 
where $\Gamma_{nR \uparrow}$ 
$\gg$ $\Gamma_{nR \downarrow}$  and 
only one polarized spin current exists. In this case, 
all equations are similarly obtained without summation 
with $\sigma$ and the factor 1/2. 

In our formulation the resonant level 
is assumed to be due to a quantum dot such as 
a small magnetic particle,  
however, the equations derived above are general forms and the 
results obtained here are 
considered to be intrinsic to the resonant tunneling 
phenomenon such as in thin film systems\cite{Liu1}.

\section{Results and Discussions}
Here we show the numerical results obtained from Eq.(\ref{ivrtc}).
The current-voltage characteristics through a double barrier 
structure reflect the 
effect of DOS of the electrodes and the magnitude 
of the Fermi surface\cite{Liu2}. 
The form of the $I$-$V$ curve has a peak when the 
Fermi energy fits the discrete energy-level 
and becomes lower as the bias voltage becomes higher 
in the three-dimensional electrode. 
The width of the peak current shows the magnitude 
of the Fermi energy when the discrete energy-level 
passes through the Fermi energy of the electrode. 
The existence of an internal magnetic field  
makes the DOS of up spins and that of down spins  
different, and the $I$-$V$ curve shows a dip reflecting 
these two different DOS at the Fermi energy. 
These features of the model are represented in Fig. \ref{fig2} 
which shows the $I$-$V$ curve and $MR=(J_{\uparrow \uparrow}
-J_{\uparrow \downarrow})/J_{\uparrow \downarrow}$ 
of a resonant tunneling structure (Eq.(\ref{ivrtc})) with 
that of a single junction (Eq.(\ref{ivsj})) 
as a function of applied bias voltage when there is 
a non-magnetic single energy-level in an island 
and the left electrode changes its internal 
magnetic field.  
In $J_{\uparrow \uparrow}$, we take $h_L$ =1.8 eV, $h_R$=2.1eV
and $J_{\uparrow \downarrow}$, $h_L$ =-1.8 eV, $h_R$=2.1eV 
for $E_F=3$eV. 
Here we assume 
$\Gamma_{\alpha \sigma} (\epsilon) =\sqrt{\epsilon/E_{F \sigma}}
 \Gamma_{\alpha 1}(\epsilon)$ 
for the energy dependence of the three-dimensional DOS  
of the electrodes on the $I$-$V$ characteristics( Fig.1). 
To be more realistic, we take 
$ \Gamma_{\alpha 1} (\epsilon)$ $=$ $\Gamma_{\alpha 0} 
C^{(\epsilon/E_F)}$ with constant $\Gamma_{\alpha 0}$ and $C$ . 
This is an extended form used in Ref. \cite{Mier2} 
and shows that the electron with higher energy has higher 
tunneling probability. 
As discussed above, the peak current is reduced more than 
 the current of the single junction, whereas the off-resonant peak 
current is greatly enhanced, that is, more than 60\%. 
We can obtain the same relation between the $MR^{(S)}$ and $MR^{(D)}$ 
even in the case where $C=1$. 


The above results can also be applied to the MR of a leak tunneling 
current via an impurity trap state in the single junction\cite{Ricco}.  
In this case our results show that the MR of the leak current 
is less than that without a trap state  and 
it is not necessary to take care of the effect of 
the leak current in the measuring process of MR. 

\section{Conclusions}  \label{sec:concl}
In conclusion, we have studied the MR of the resonant tunneling 
through a double-barrier system by using the Keldysh formulation.  
Although our model is simpler than that of the S-matrix 
method by Zhang {\it et al.}\cite{Zhang}, it can describe the 
intrinsic characteristics of the resonant tunneling current 
which the S-matrix theory cannot treat easily.  
We showed that the peak current is not 
enhanced by the change of the magnetization 
of the soft magnet whereas 
the valley current is greatly enhanced 
when compared with the current in a single junction.  
We also found that the PV ratio decreases when 
the junction conductance increases.  

T. T. is grateful to  K. Sato  
of Toshiba Corp. for support throughout this work, 
and expresses thanks to K. Inomata, K. Mizushima, S. N. Okuno, 
Y. Saito, and T. Yamauchi for fruitful discussion.


\begin{figure}
\caption{(a) Schematic band-edge diagram for a double barrier 
structure. At sufficiently low temperature, the tunneling of 
electrons takes place 
when the discrete energy-level passes through 
the corresponding Fermi energy band. 
(b) Density of state of the up spin electron and down spin 
electrons at the electrode $\alpha$($\alpha=L,R$). $h_\alpha$ 
is an internal magnetic field. }
\label{fig1}
\end{figure}

\begin{figure}
\caption{Resonant tunneling current and the MR (see text) 
in the case where the left electrode is a soft magnet. 
$MR^{(D)}$ is the MR of the double barrier resonant tunneling 
current and $MR^{(S)}$ is that of the single barrier. 
In $J_{\uparrow \uparrow}$, $h_L/E_F=0.6$ and $h_R/E_F=0.7$ and 
in $J_{\uparrow \downarrow}$, $h_L/E_F=-0.6$ and $h_R/E_F=0.7$ 
for $E_{F}$= 3.0eV where $T/E_F=10^{-3}$, 
$\Gamma_{L0}/E_F$=$5.0\times 10^{-5}$, 
$\Gamma_{R0}/E_F$=$2.5\times 10^{-5}$ and $C=10$  (see text). 
The $MR^{(D)}$ and $MR^{(S)}$ refer to the right scale and 
the $J_{\uparrow \uparrow}$ and $J_{\uparrow \downarrow}$ 
refer to the left one.
$MR^{(D)}$ at the peak current is smaller than $MR^{(S)}$, 
whereas that at the valley current is enhanced more than 60\%. }
\label{fig2}
\end{figure}



\begin{references}
\bibitem{Johnson}
M. Johnson :Science {\bf 260}, 320 (1993).

\bibitem{Moodera}
J. S. Moodera and L. R. Kinder: J. Appl. Phys. {\bf 79}, 4724 (1996), 
and references therein. 

\bibitem{Maekawa}
S. Maekawa and U. G$\ddot{\rm a}$fvert: 
IEEE Trans. Magn. {\bf MAG-18}, 707 (1982).

\bibitem{Slonczewski}
J. C. Slonczewski:Phys. Rev. B {\bf 39}, 6995 (1989).

\bibitem{Zhang}
X. Zhang, B. Li, G. Sun and F. Pu: 
Phys. Rev. B {\bf 56}, 5484 (1997).

\bibitem{Jauho}
A. P. Jauho, N. S. Wingreen and Y. Mier, 
Phys. Rev. B {\bf 50}, 5528 (1994). 

\bibitem{Mier}
Y. Mier and N. S. Wingreen, 
Phys. Rev. Lett. {\bf 68}, 2512 (1992). 

\bibitem{Sarker}
S. K. Sarker: Phys. Rev. {\bf B32}, 743 (1985).

\bibitem{Weil}
T. Weil and B. Vinter, 
Appl. Phys. Lett. {\bf 50}, 1281 (1987); 
M. Jonson and A. Grincwaig, {\it ibid} {\bf 51}, 1729 (1987)

\bibitem{Liu1}
H. C. Liu, M. Buchanan, G. C. Aers, Z. R. Wasilewski, 
W. T. Moore, R. L. S. Devine and D. Landheer: 
Phys. Rev. {\bf B 43} , 7086 (1991).

\bibitem{Liu2}
H. C. Liu and G. C. Aers: Appl. Phys. Lett. {\bf 65}, 4908 (1989). 

\bibitem{Mier2}
Y. Mier, N. S. Wingreen, and P. A. Lee, 
Phys. Rev. Lett. {\bf 66}, 3048 (1991). 

\bibitem{Ricco}
B. Ricco and M. Ya. Azbel and M. H. Brodsky: 
 Phys. Rev. Lett. {\bf 51}, 1795 (1983).

\end{references}
\end{document}